\documentclass[a4paper,12pt]{article}
\usepackage{color}
\usepackage{epsfig}
\usepackage{amsfonts}
\usepackage{amsmath}
\usepackage{amssymb, amsbsy}
\usepackage{mathtools}
\usepackage{chngpage, setspace}
\usepackage{lineno}
\usepackage{url}
\usepackage{pdfpages}
\begin{document}
\title{Context-stratified Mendelian randomization: exploiting regional exposure variation to explore causal effect heterogeneity and non-linearity}
\author{Stephen Burgess \textsuperscript{1,2} \and Benjamin A R Woolf \textsuperscript{1,3,4} \and Amy M Mason \textsuperscript{2,5}}
\date{}
\maketitle

\vspace{3mm}

\noindent \noindent \textsuperscript{1} Medical Research Council Biostatistics Unit, University of Cambridge, Cambridge, UK. \vspace{2mm} \\
\textsuperscript{2} British Heart Foundation Cardiovascular Epidemiology Unit, Department of Public Health and Primary Care, University of Cambridge, Cambridge, UK. \vspace{2mm} \\
\textsuperscript{3} Medical Research Council Integrative Epidemiology Unit, University of Bristol, Bristol, UK. \vspace{2mm} \\
\textsuperscript{4} School of Psychological Science, University of Bristol, Bristol, UK. \vspace{2mm} \\
\textsuperscript{5} Victor Phillip Dahdaleh Heart and Lung Research Institute, University of Cambridge, Cambridge, UK.

\vspace{3mm}

\noindent \noindent \textbf{Corresponding author:} \\ \indent Stephen Burgess, \\ \indent MRC Biostatistics Unit, \\ \indent University of Cambridge, \\ \indent East Forvie Building, \\ \indent Robinson Way, \\ \indent Cambridge, Cambridgeshire, CB2 0SR, UK. \\ \indent Tel: +44 (0) 1223 768259. \\ \indent Email: sb452@medschl.cam.ac.uk.

\vspace{3mm}

\noindent \noindent \textbf{Funding:} SB and BARW are supported by the Wellcome Trust (225790/Z/22/Z) and the United Kingdom Research and Innovation Medical Research Council (MC\_UU\_00040/01). This work was supported by core funding from the British Heart Foundation (RG/F/23/110103), NIHR Cambridge Biomedical Research Centre (NIHR203312) [*], BHF Chair Award (CH/12/2/29428), and by Health Data Research UK, which is funded by the UK Medical Research Council, Engineering and Physical Sciences Research Council, Economic and Social Research Council, Department of Health and Social Care (England), Chief Scientist Office of the Scottish Government Health and Social Care Directorates, Health and Social Care Research and Development Division (Welsh Government), Public Health Agency (Northern Ireland), British Heart Foundation and the Wellcome Trust.

\clearpage

\section*{Abstract}
Mendelian randomization (MR) uses genetic variants as instrumental variables to make causal claims. Standard MR approaches typically report a single population-averaged estimate, limiting their ability to explore effect heterogeneity or non-linear dose--response relationships. Existing stratification methods, such as residual-based and doubly-ranked stratified MR, attempt to overcome this but rely on strong and unverifiable assumptions. We propose an alternative, context-stratified Mendelian randomization, which exploits exogenous variation in the exposure across subgroups -- such as recruitment centres, geographic regions, or time periods -- to investigate effect heterogeneity and non-linearity. Separate MR analyses are performed within each context, and heterogeneity in the resulting estimates is assessed using Cochran's Q statistic and meta-regression.

We demonstrate through simulations that the approach detects heterogeneity when present while maintaining nominal false positive rates under homogeneity when appropriate methods are used. In an applied example using UK Biobank data, we assess the effect of vitamin D levels on coronary artery disease risk across 20 recruitment centres. Despite some regional variation in vitamin D distributions, there is no evidence for a causal effect or heterogeneity in estimates. Compared to stratification methods requiring model-based assumptions, the context-stratified approach is simple to implement and robust to collider bias, provided the context variable is exogenous. However, the method's power and interpretability depend critically on meaningful exogenous variation in exposure distributions between contexts. In the example of vitamin D, subgroups from other stratification methods explored a much wider range of the exposure distribution.

\vspace{5mm}

\noindent \noindent \textbf{Keywords:} instrumental variables, effect heterogeneity, meta-regression, non-linearity, stratification, causal inference.

\clearpage

\section{Introduction}
Mendelian randomization uses genetic variants as instrumental variables (IVs) to estimate the causal effect of a modifiable exposure on an outcome \cite{lawlor2007, burgess2021book}. Under the core IV assumptions (relevance, independence, and exclusion restriction), Mendelian randomization can provide unbiased causal estimates by leveraging randomness in the inheritance of genetic variants at conception as a natural experiment \cite{didelez2007}. Conventional Mendelian randomization typically reports an average causal effect estimate for the population \cite{angrist1996}. This is analogous to the average treatment effect in a randomized controlled trial, which is averaged over all individuals whose exposure value is altered by randomization \cite{imbens1994}.

However, causal effects may not be constant across all individuals or settings. In many cases, one may expect effect heterogeneity: the exposure might have a stronger effect in some subgroups of the population than others \cite{tian2024, man2025}. Alternatively, a non-linear dose-response relationship might exist \cite{burgess2014nonlin}. In a randomized controlled trial, investigators often explore this by performing stratified analyses (also called subgroup analyses) to see if the treatment effect differs by baseline risk factors or other characteristics \cite{wang2007}. Identifying such heterogeneity is critical for targeting interventions to those who would benefit most. For example, if an exposure is only causally beneficial when baseline levels are low (a threshold effect), a population-averaged estimate would fail to identify this important nuance.

Performing subgroup analyses in Mendelian randomization is challenging. Any covariate measured after conception is technically ``post-randomization'' (since genetic randomization occurs at conception). Naively stratifying a Mendelian randomization analysis on a post-randomization covariate (especially one affected by the genotype or exposure) can induce collider bias \cite{cole2010}, potentially yielding biased stratum-specific estimates \cite{yusuf1991}. For instance, stratifying on the exposure level itself (or a variable influenced by it) will induce correlations between the instrument and confounders within strata, invalidating estimates \cite{burgess2022crp}. Exceptions to this rule are covariates such as age, sex, and ancestry, which are either determined by external factors or fixed from conception.

Recent methodological work has highlighted these concerns and proposed solutions for safer stratification in Mendelian randomization. In particular, the residual-based and doubly-ranked stratification methods have been developed to form exposure strata while preserving instrument validity \cite{staley2017, tian2022relaxing}. Under specific assumptions, these techniques ensure that genetic instruments remain valid instruments within each stratum, avoiding bias when exploring non-linear effects. However, the assumptions that these methods make go beyond the standard IV assumptions, and violations of these assumptions can lead to bias, and potentially to misleading findings \cite{burgess2023violation, hamilton2025}.

An alternative proposal is to investigate how the causal effect varies across subgroups that differ in their distribution of the exposure, but are defined in a way that is not based on the genetic variants or the exposure. A natural way to do this is to use structure within the data, such as recruitment centres, which we refer to as contexts. The central idea is simple: if the average level of the exposure differs between subgroups, then performing Mendelian randomization within each subgroup provides a causal estimate relevant to that exposure distribution. Comparing these estimates can reveal effect heterogeneity or non-linearity. We refer to this approach as ``context-stratified Mendelian randomization''.

This approach treats context as analogous to a baseline effect modifier. It is conceptually akin to a multi-site trial analyzing treatment effects separately at each site to see if they differ by some site-level condition \cite{feaster2011}. It is similarly analogous to meta-regression, in which study-specific estimates are regressed on a study covariate to investigate trends in estimates \cite{thompson2002}. We note that while estimates in a randomized controlled trial (both in main analyses and in subgroup analyses stratified by baseline covariates) are typically unconfounded due to randomization, both comparisons between trial estimates and comparisons of subgroup estimates are typically confounded.

In this paper, we first describe the methodology of context-stratified Mendelian randomization (Section~2). We demonstrate through simulations that the approach can successfully identify non-linear causal patterns and does not detect false heterogeneity when the true effect is homogeneous and appropriate methods are used (Section~3). We present an example of its application in UK Biobank, evaluating estimates of the effect of vitamin D levels on coronary artery disease risk in different recruitment centres, and assessing heterogeneity and trends in estimates across centres (Section~4). We then discuss the insights gained, the differences from other approaches, and the potential implications for epidemiological research and targeted public health interventions (Section~5). In particular, we discuss the interpretation of differences in estimates between contexts.

\section{Methods}
\subsection{Concept}
Context-stratified Mendelian randomization involves splitting the study population into subgroups based on an external context variable that influences the exposure distribution. Typical context variables could be study regions, recruitment centres, time periods, age groups, or other pre-specified subpopulations. The key is that these subgroups have different average levels of the exposure of interest, and that the context variable is exogenous; that is, it is not a function of any variable in the model. We then perform separate Mendelian randomization analyses within each context to obtain context-specific causal estimates. By examining how these estimates differ across contexts, we can infer whether the causal effect might depend on the exposure distribution (suggesting effect modification or non-linearity).

\subsection{Assumptions}
Within each subgroup defined by the context variable, the standard IV assumptions must hold for any genetic variant used as an instrument:
\begin{enumerate}
  \item \textbf{Relevance:} The genetic variant is associated with the exposure in each subgroup.
  \item \textbf{Independence:} There is no confounding pathway between the variant and outcome in any subgroup.
  \item \textbf{Exclusion restriction:} The variant can only affect the outcome through the exposure.
\end{enumerate}

Because we stratify on a context variable that is external to the variables in the model, this stratification should not introduce collider bias. As analyses are performed separately in each subgroup, it is not necessary for the genetic variants to be distributed identically in the subgroups. However, if we want to attribute differences in context-specific estimates to the exposure, then other factors should be comparable across subgroups.

\subsection{Analytical steps}
The implementation of context-stratified Mendelian randomization can be summarized as follows:

\begin{itemize}
\item Step 1: Select a context variable that partitions the dataset into $K$ subgroups with distinct exposure distributions.
\item Step 2: For each subgroup $k$, calculate a summarized measure of the exposure (for example, the mean or median exposure level). These will serve as indicators of the exposure distribution for interpretation and for meta-regression modelling.
\item Step 3: Within each subgroup $k$, conduct an IV analysis to estimate the causal effect of the exposure on the outcome. With a continuous outcome, this could be done via two-stage least squares if individual-level data are available or via the ratio method if a single genetic instrument is used. Standard summarized-data methods such as the inverse-variance weighted method could be used if context-specific summarized data are available \cite{burgess2015scoretj}.
\item Step 4: Compare the context-specific IV estimates $\hat{\beta}_1, \hat{\beta}_2, \ldots, \hat{\beta}_K$. A formal heterogeneity test can be performed using Cochran's $Q$ statistic \cite{bowden2018invited}. A significant $Q$ statistic would indicate that the estimates are more variable than expected by chance, consistent with effect heterogeneity or non-linearity.
\item Step 5: To quantify and visualize the relationship between causal estimates and the exposure distribution, perform a meta-regression. In a meta-regression, the dependent variable is the vector of context-specific IV estimates, and the key independent variable is the vector of average exposure levels in each subgroup. A positive slope, for instance, would indicate contexts with larger mean exposure levels have greater causal estimates (or vice versa). The trend can also be explored visually by a scatter plot of the context-specific IV estimates against the mean exposure values.
\end{itemize}

\section{Simulation study}
\subsection{Set-up}
To validate the context-stratified Mendelian randomization approach, we conducted a simulation study. We simulate a continuous exposure $X$, a genetic instrument $G$ (taking values 0,1,2 to mimic an allele count), an unmeasured confounder $U$, and a continuous outcome $Y$. We generate data on 10\thinspace000 individuals in each of 10 contexts, making a total of 100\thinspace000 individuals. The mean level of the exposure is varied between contexts by adding a constant $\alpha_k$. We consider two scenarios for between-context differences: a larger difference scenario where $\alpha_k$ is 8 in the first context, 8.2 in the second context, then 8.4 and so on up to 9.8, and a smaller difference scenario where $\alpha_k$ is 9 in the first context, 9.1 in the second context, then 9.2 and so on up to 9.9. We consider three scenarios for the true causal effect of the exposure on the outcome:
\begin{enumerate}
  \item linear homogeneous effect (no effect modification or non-linearity): $f(x) = 0.8 x$.
  \item quadratic effect: $f(x) = 0.04 x^2$.
  \item threshold effect: $f(x) = 0.25 \, I(x>10) \, (x-10)$
\end{enumerate}
where $I(x>10)$ is an indicator function that is 1 if the exposure is greater than 10, and 0 otherwise.

The data-generating model for individual $i$ in context $k$ is as follows:
\begin{align}\label{eq:datgen}
  g_{ik} &\sim \mbox{Binomial}(2, 0.3) \notag \\
  u_{ik}, \epsilon_{Xik}, \epsilon_{Yik} &\sim \mathcal{N}(0, 1) \mbox{ independently} \notag \\
  x_{ik} &= \alpha_k + 0.5 g_{ik} + u_{ik} + \epsilon_{Xik} \notag \\
  y_{ik} &= f(x_{ik}) - u_{ik} + \epsilon_{Yik} \notag
\end{align}

We generate 1000 simulated datasets for each of the six scenarios. For each simulated dataset, we calculate the $k$th context-specific IV estimate using the ratio method as the genetic association with the outcome in context $k$ divided by the genetic association with the exposure in context $k$. We then calculate Cochran's $Q$ statistic as a measure of heterogeneity in the context-specific IV estimates, and perform meta-regression of the context-specific IV estimates on the mean level of the exposure in each context.

Various versions of Cochran's $Q$ statistic have been proposed, as the method relies on the standard errors of the estimates to use as weights, and there are different ways of calculating these standard errors. We calculate Cochran's $Q$ statistic in two ways: using first-order weights, and using modified second-order weights \cite{bowden2017weights}. Code to run the simulation study is provided in the Supplementary Material.

\subsection{Results}
Results from the simulation study are shown in Table~\ref{tab:simres}. For each scenario, we report the proportion of simulated datasets in which heterogeneity was detected (defined as the p-value for the $Q$ statistic being less than 0.05), and the proportion of simulated datasets in which a trend was detected (defined as the p-value for the mean exposure term in the meta-regression being less than 0.05). The genetic instrument explained around 4.3\% of variance in the exposure (larger differences) or 4.8\% of variance (smaller differences), corresponding to an average F statistic of around 4500.

With larger differences, in the linear scenario, the proportion of simulated datasets in which heterogeneity is detected is 12.8\% when using first-order weights. This is greater than the expected 5\% rate that we would expect under the null hypothesis. A similar finding has been observed previously, in which the heterogeneity test over-rejects the null even with conventionally strong instruments \cite{verbanck2018}. When the heterogeneity test is being used to detect pleiotropy, over-rejection of the test is tolerable, as between-variant differences in estimates are interpreted as evidence suggesting pleiotropy. However, in this case, we do not want to over-reject the null hypothesis. When using modified second-order weights, the proportion of simulated datasets in which heterogeneity is detected is 0.4\%, well below the nominal 5\% level. The proportion of datasets in which a trend is detected when the true effect is linear is 3.3\%, close to the expected 5\% level.

In contrast, in the quadratic and threshold scenarios, the proportion of simulated datasets rejecting the null is well over 5\% for both versions of the heterogeneity test, and for the trend test. We do not interpret these numbers too strongly, as power to detect differences in estimates depends on many factors, including sample size, the difference in the exposure distribution between subgroups, and the instrument strength. However, we note that this a fairly optimistic setting, in which the instrument is strong, the sample size is large, and differences in the exposure distribution between subgroups are substantial. Even in this setting, power to detect differences between estimates is far from perfect.

With smaller differences, results are similar in the linear scenario, with elevated coverage rates for the heterogeneity test using first-order weights of 10.4\%, conservative rates for the heterogeneity test using modified second-order weights of 0.1\%, and close to nominal levels for the trend test of 4.1\%. In the quadratic and threshold scenarios, the proportion of simulated datasets rejecting the null is much lower than with larger differences, and even was below 5\% with the quadratic model for the heterogeneity test using modified second-order weights.

While using modified second-order weights in calculation of the $Q$ statistic is necessary to avoid over-rejection of the null, estimates of the $Q$ statistic with modified second-order weights for moderate strength instruments are often lower than with first order weights, and hence power to detect true heterogeneity is lower. Users may face a pragmatic choice as to which version of the heterogeneity method to use: whether to use the version with non-inflated coverage properties under the null, but lower power (modified second-order); or the version with better power but inflated coverage under the null (first-order). The trend test assumes that there is a linear trend between estimates and mean exposure levels in the subgroups, which would not hold if the causal model is U-shaped. However, if the hypothesized causal model is monotone in the mean exposure levels (that is, the causal effect across the exposure distribution is everywhere either zero or positive, or it is everywhere either zero or negative), then the trend test will typically have greater power to detect differences in estimates, as was seen in our simulation study.

\begin{table}[hbtp]
\begin{center}
\centering
\begin{tabular}[c]{c|ccc}
\hline
Scenario  & Heterogeneity detected? & Heterogeneity detected? & Trend detected? \\
          & (first-order weights)   & (modified second-order) &                 \\
\hline
\multicolumn{4}{c}{Larger differences between subgroups}                        \\
\hline
Linear    & 12.8\%                  &  0.4\%                  &  3.3\%          \\
Quadratic & 76.4\%                  & 28.4\%                  & 88.7\%          \\
Threshold & 42.3\%                  & 41.7\%                  & 64.9\%          \\
\hline
\multicolumn{4}{c}{Smaller differences between subgroups}                       \\
\hline
Linear    & 10.4\%                  &  0.1\%                  &  4.1\%          \\
Quadratic & 26.7\%                  &  3.3\%                  & 37.6\%          \\
Threshold & 18.1\%                  & 16.8\%                  & 26.5\%          \\
\hline
\end{tabular}
\caption{Simulation study results: proportion of simulated datasets in which heterogeneity was detected (defined as the p-value for the $Q$ statistic being less than 0.05, using first-order and modified second-order weights) and proportion in which a trend was detected (defined as the p-value for the mean exposure term in the meta-regression being less than 0.05) with three different models for the effect of the exposure on the outcome and two sets of values of the differences between groups.}
\label{tab:simres}
\end{center}
\end{table}

\section{Applied example}
We illustrate use of the context-stratified Mendelian randomization approach in an applied example. We perform separate Mendelian randomization analyses in recruitment centres from the UK Biobank to investigate the effect of 25-hydroxyvitamin D [25(OH)D], a circulating metabolite indicating vitamin D levels, on coronary artery disease risk. We selected 25(OH)D as our exposure of interest as its levels depend greatly on environmental context, and on sunlight exposure in particular.

The UK Biobank is a large prospective cohort of around 500\thinspace000 participants aged 40 to 69 years at baseline, recruited across 22 assessment centres throughout the UK from 2006 to 2010. Our analysis includes data from 20 centres, excluding Stockport (which was the pilot centre) and Wrexham (which had much lower participant numbers than other centres). In total, we considered data on 323\thinspace539 unrelated European-ancestry individuals with a valid 25(OH)D measurement using the same exclusion and quality control filters as in a previous paper \cite{sofianopoulou2024}. We use a genetic instrument that is a weighted score comprising 21 variants located in 4 gene regions that have specific biological relevance to vitamin D transport, metabolism, and synthesis, as previously used in the same paper. The exposure was shifted to account for the month of blood draw in a similar way to the correction for season of blood draw in the previous paper; values represent a measurement taken in October. The outcome was also defined in the same way as in this paper, using International Classification of Diseases, Tenth Revision (ICD-10) codes for fatal ischaemic heart disease (ICD-10 code: I20-I25) or non-fatal myocardial infarction (I21-I23).

Estimates were obtained using the ratio method with a weighted genetic score as the instrument. Genetic associations with 25(OH)D levels were estimated using linear regression. Genetic associations with coronary artery disease were estimated using logistic regression. In all regression models, we adjust for age at baseline, sex, and 10 genomic principal components of ancestry.

Details of the participants are provided in Table~\ref{tab:baseline}. We see that mean 25(OH)D levels vary somewhat across contexts, with lowest levels in the two Scottish recruitment centres, and highest levels in the two Welsh recruitment centres. The difference in mean values from the lowest to the highest centre is 7.2 nmol/L, which is around 0.4 standard deviations. In contrast, mean levels of body mass index (BMI), systolic blood pressure (SBP), and low-density lipoprotein cholesterol (LDL-c) do not differ as greatly between the centres. This is reassuring for our analysis, as we want to attribute any differences in context-specific estimates to 25(OH)D levels. However, it means that similar analyses will be difficult for other exposures, particularly for LDL-c.

\begin{table}[hbtp]
\begin{center}
\centering
\begin{tabular}[c]{c|cccccc}
\hline
Recruitment    & Sample size &   Age   &  Vit D   &  BMI       &  SBP   &  LDL-c   \\
centre         &     (n)     & (years) & (nmol/L) & (kg/m$^2$) & (mmHg) & (mmol/L) \\
\hline
Glasgow        &    13501    &   56.7  &  50.2    &  27.6  &  139.0  &  3.49  \\
Edinburgh      &    12258    &   56.7  &  50.9    &  27.0  &  138.5  &  3.54  \\
Barts          &     7563    &   55.6  &  52.5    &  26.2  &  131.0  &  3.44  \\
Newcastle      &    23819    &   57.2  &  54.0    &  27.7  &  138.5  &  3.50  \\
Manchester     &     9502    &   56.0  &  54.0    &  27.4  &  137.2  &  3.44  \\
Birmingham     &    15779    &   57.4  &  54.9    &  27.7  &  137.4  &  3.45  \\
Leeds          &    29509    &   57.1  &  55.0    &  27.4  &  137.8  &  3.50  \\
Croydon        &    16688    &   57.6  &  55.2    &  27.0  &  134.8  &  3.47  \\
Stoke          &    12843    &   57.0  &  55.4    &  27.7  &  139.1  &  3.49  \\
Bury           &    19993    &   57.7  &  55.5    &  27.7  &  137.9  &  3.45  \\
Oxford         &    10499    &   56.8  &  55.8    &  26.6  &  135.4  &  3.52  \\
Nottingham     &    22831    &   57.5  &  56.1    &  27.4  &  137.4  &  3.50  \\
Middlesborough &    13932    &   57.4  &  56.1    &  27.8  &  141.0  &  3.53  \\
Liverpool      &    21479    &   57.8  &  56.3    &  27.8  &  137.6  &  3.44  \\
Hounslow       &    17276    &   57.1  &  56.4    &  26.7  &  134.6  &  3.43  \\
Sheffield      &    19748    &   57.7  &  56.8    &  27.5  &  138.1  &  3.50  \\
Reading        &    21526    &   57.0  &  56.9    &  26.8  &  134.9  &  3.53  \\
Bristol        &    29527    &   56.3  &  56.9    &  27.0  &  138.8  &  3.52  \\
Swansea        &     1562    &   58.1  &  57.1    &  28.1  &  141.6  &  3.51  \\
Cardiff        &    12705    &   56.4  &  57.4    &  27.9  &  138.5  &  3.49  \\
\hline
\end{tabular}
\caption{Baseline characteristics of UK Biobank participants presented as centre-specific means. Centres are ordered based on mean 25-hydroxyvitamin D [25(OH)D] levels from lowest to highest. For each centre, we list the total number of eligible participants (sample size, n), mean age (years), mean 25(OH)D levels (Vit D, nmol/L), mean body mass index (BMI, kg/m$^2$), mean systolic blood pressure (SBP, mmHg), and mean low-density lipoprotein cholesterol (LDL-c, mmol/L) levels.}
\label{tab:baseline}
\end{center}
\end{table}

\begin{figure}[htbp]
    \centering
 \includegraphics[width=0.49\textwidth]{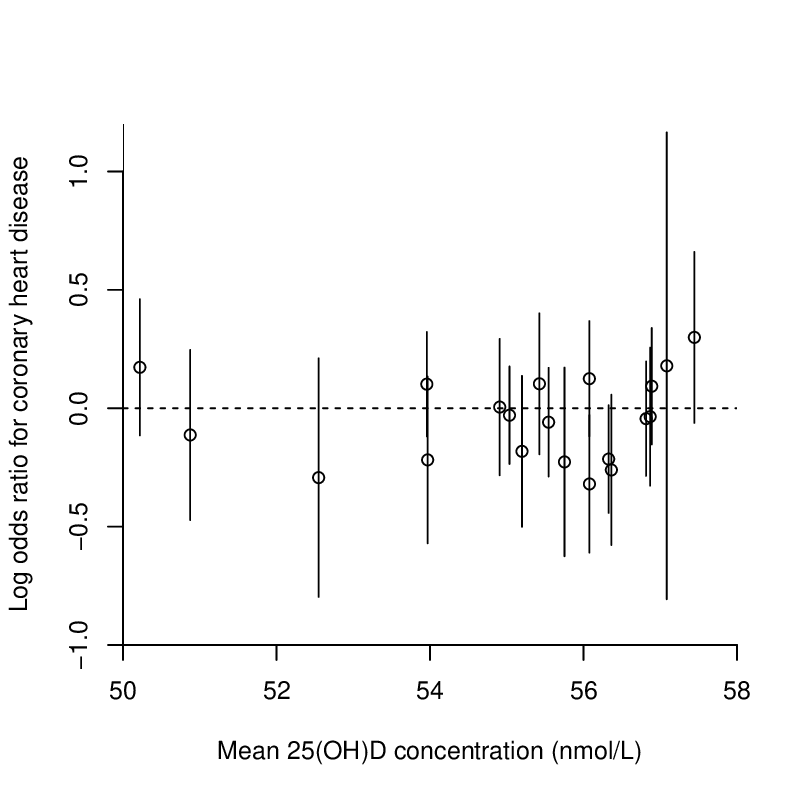}
 \includegraphics[width=0.49\textwidth]{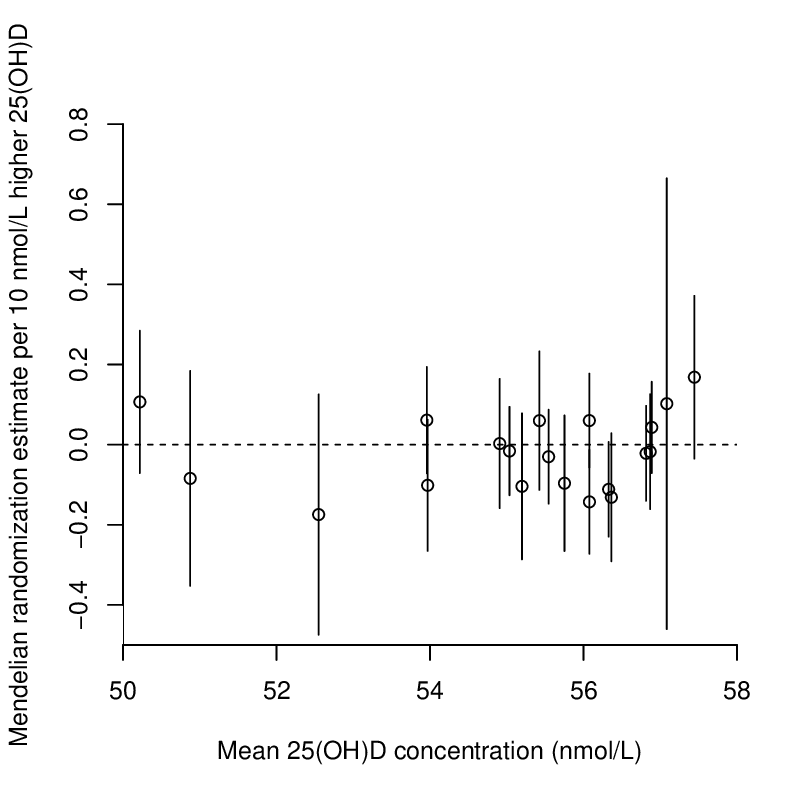}
\caption{Scatter plots of context-specific estimates (95\% confidence intervals) against mean 25-hydroxyvitamin D in each subgroup. Left panel: association of weighted genetic score with coronary artery disease risk (log odds ratio). Right panel: Mendelian randomization estimates scaled to 10 nmol/L higher genetically-predicted 25(OH)D levels.} \label{fig:scatter}
\end{figure}

Scatter plots of context-specific estimates against mean exposure levels are provided in Figure~\ref{fig:scatter}. In Figure~\ref{fig:scatter} (left panel), we provide the unscaled associations of the weighted genetic score with coronary artery disease risk. In Figure~\ref{fig:scatter} (right panel), we provide the Mendelian randomization estimates scaled to 10 nmol/L higher genetically-predicted 25(OH)D levels. Estimates in all subgroups are compatible with the null, and there is no evidence of differences in estimates: heterogeneity $Q$ statistic (first-order) = 22.2, $p = 0.28$; heterogeneity $Q$ statistic (modified second-order) = 22.2, $p = 0.28$; trend test $p = 0.76$.

\begin{figure}[htbp]
    \centering
 \includegraphics[width=0.99\textwidth]{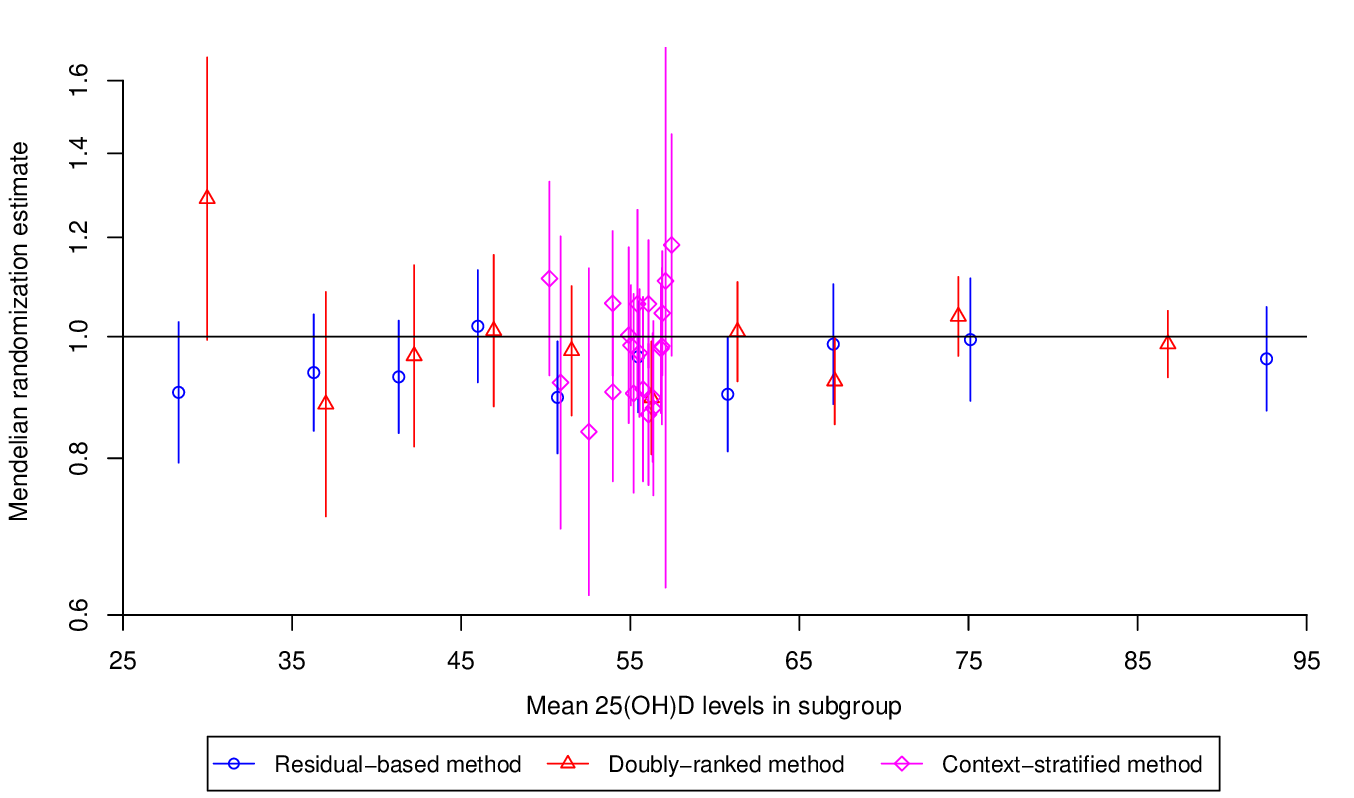}
\caption{Scatter plots of Mendelian randomization estimates (95\% confidence intervals) scaled to 10 nmol/L higher genetically-predicted 25(OH)D levels against mean 25-hydroxyvitamin D for subgroups defined by the residual-based, doubly-ranked, and context-stratified methods.} \label{fig:compare}
\end{figure}

We have previously considered estimates of the effect of 25(OH)D on coronary artery disease risk in 10 subgroups of the population defined using the residual-based and doubly-ranked methods \cite{burgess2023violation, sofianopoulou2024}. For comparison, we plot our context-specific estimates on the same axes as these stratum-specific estimates in Figure~\ref{fig:compare}. We see that estimates from the stratified methods cover a much wider range than those from the context-stratified method. While the mean 25(OH)D levels in the strata from the context-stratified method are all between 50 and 58 nmol/L, mean levels in strata defined by the residual-based and doubly-ranked methods range from below 30 to above 85 nmol/L.

\section{Discussion}
In this paper, we have presented context-stratified Mendelian randomization as an approach to assess causal effect heterogeneity using Mendelian randomization, and illustrated its use through simulations and an applied example. This approach is conceptually straightforward yet potentially powerful for detecting variability in causal effects across subpopulations defined using an exogenous variable, such as recruitment centre. A simulation study showed that the approach can reliably detect non-linearity, particularly using a trend test in meta-regression. We illustrated the approach using data from UK Biobank. While we were not able to detect evidence for a causal effect, let alone heterogeneity in causal estimates, this applied example illustrates the feasibility of the approach for exploring differences in estimates.

\subsection{Comparison with previous methods}
Several methods have been proposed for investigating effect heterogeneity and non-linearity using instrumental variables \cite{burgess2024towards}. Approaches using non-linear least squares are attractive in some settings \cite{amemiya1974}, but in Mendelian randomization where the IVs typically only explain a small proportion of variance in the exposure, fitted values of the exposure from the first-stage regression only span a small section of the exposure distribution, meaning that findings from such methods are at best limited in scope. Such approaches can also be sensitive to the specification of the non-linear model \cite{mogstad2010}. While flexible non-parametric alternatives have been developed \cite{hartford2017deep, singh2019kernel, bennett2019deep}, any estimates beyond the exposure values predicted by the IVs rely on blind extrapolation. More promising are approaches that use the residuals from the first-stage model \cite{guo2016control, sulc2022}, although again these approaches can be sensitive to parametric assumptions and model specification choices.

Alternatively, several stratification approaches have been proposed, which are more similar to the context-stratified Mendelian randomization approach discussed here. The residual-based stratification method calculates residuals from regression of the exposure on the IVs, which represent the value that the exposure would take at a fixed level of the IVs \cite{burgess2014nonlin, staley2017}. Stratification on these residual values allows estimates to be calculated for subgroups with large differences in their mean exposure levels. However, the residual-based stratification method makes a strong and unrealistic assumption (known as the constant-effect assumption) that the effect of the genetic variants is constant in the population \cite{hamilton2024one}. The doubly-ranked stratification method is more flexible, stratifying individuals based on their percentile of the counterfactual exposure distribution at a given value of the IVs, under the assumption (known as the rank-preserving assumption) that this percentile would be the same at all values of the IVs \cite{tian2022relaxing}. This is a strictly weaker assumption than the constant-effect assumption, and is more realistic. However, subgroups defined by the doubly-ranked method are slightly more similar as to their average levels of the exposure compared to those from the residual-based method. Additionally, violation of the rank-preserving assumption would lead to bias \cite{burgess2023violation, hamilton2025}.

The context-stratified Mendelian randomization approach makes a more justifiable assumption, relying on the existence of subgroups defined by a natural division of the study population. However, in the example of vitamin D considered in this paper, subgroups defined by the context-stratified approach are far more similar as to their average levels of the exposure than those from other stratified methods. We chose to investigate vitamin D as an exposure as this has plausible variation in levels between centres. However, even for this example, variability in context-specific mean exposure levels was limited. For other exposures, it may not be possible to find a natural division into subgroups to facilitate this approach.

A similar context-stratified approach was previously performed to investigate the effect of alcohol consumption in different geographic regions of China \cite{millwood2019}. However, rather than performing separate analyses in each region, analysts defined a composite instrument score based on genetic variants and geographic region. While they did adjust for geographic region in primary analyses, defining an IV using geographic region could conflate genetic differences and regional differences, such that comparisons are not solely driven by differences in genetic variants.

Other related approaches in Mendelian randomization have considered the comparison of estimates in different subgroups, but typically the focus of these approaches has been to detect or avoid pleiotropic bias rather than to explore effect heterogeneity or non-linearity. Subgroups in whom the exposure is absent (such as women in analyses investigating the effect of alcohol consumption in East Asia, where women do not tend to drink alcohol for cultural reasons \cite{lewis2005b}) can serve as natural negative control populations \cite{vankippersluis2017}. Gene-by-environment interactions can be exploited to account for pleiotropy under certain assumptions about the consistency of pleiotropic effects across subgroups \cite{spiller2019}. Additionally, the MR-GENIUS (Mendelian randomization G-estimation under No Interaction with Unmeasured Selection) method leverages variance differences in the exposure distribution across genotype subgroups (implicitly due to some unknown interactions) to obtain a single robust estimate \cite{tchetgen2021}. In other work, associations of the lead variant in the \emph{FTO} gene region with BMI have been shown to vary with age \cite{pagoni2024}, and some differences in stratified Mendelian randomization estimates defined according to age and urate levels have been observed \cite{man2025}.

\subsection{Interpretation of results}
In a randomized controlled trial, differences in subgroup analysis estimates are not necessarily attributable (or not fully attributable) to the stratification trait, as this is not a randomized comparison. For instance, estimates may be higher in men than in women due to greater mean levels of BMI in men, rather than due to any aspect of sex or gender. Similarly in stratified non-linear Mendelian randomization using the residual-based or doubly-ranked method: differences between estimates are not necessarily fully attributable to the exposure, but may be attributable to differences in the strata populations \cite{small2014}. A similar statement can be made about context-stratified Mendelian randomization, except that in this approach, subgroups are primarily defined by the context variable. Hence we may expect these differences to be even more marked in context-stratified Mendelian randomization. Equally, differences in the exposure distribution between subgroups may be less marked. In the applied example in this paper, there is an intrinsic correlation between average exposure levels and geographical location, and hence any differences in estimates may be due to factors that also vary by geographical location. We would urge caution in the interpretation of findings where differences between estimates are detected.

\subsection{Limitations}
Compared with other stratification approaches for non-linear Mendelian randomization, context-stratified Mendelian randomization is relatively intuitive. As it does not rely on a technical assumption, but rather exogeneity of the context variable, context-specific estimates are more likely to be valid causal estimates than stratum-specific estimates from the residual-based or doubly-ranked method, in which inferences depend on more questionable statistical assumptions. However, finding a relevant context variable is likely to be challenging for exposures that do not differ systematically in their distribution across the population. Even for vitamin D, differences in mean levels between UK Biobank centres were not substantial. Additionally, differences between estimates in context-stratified Mendelian randomization are more likely to be explained by another variable other than the exposure than estimates from other stratification approaches. Further, as with all methods for making causal inferences, conclusions depend on assumptions that are untestable. When we have a feasible context variable, the assumptions for context-stratified Mendelian randomization methods are more reasonable than those for other non-linear Mendelian randomization methods for any given exposure. However, all approaches depend on the validity of the genetic variants as instrumental variables.

\subsection{Conclusions}
In conclusion, context-stratified Mendelian randomization provides an accessible and conceptually appealing addition to the epidemiologist's toolbox. By exploiting contextual variability in the distribution of the exposure, it allows us to ask questions about effect heterogeneity and non-linearity. In examples where a relevant context variable can be found, it is a worthwhile exploratory approach to investigate differences in estimates that may be attributable to heterogeneity or non-linearity in the effect of an exposure. However, differences between estimates may not be attributable to the exposure. Additionally, even in cases where it is feasible, the range of exposure levels in strata defined by the method may be far narrower than those defined by other stratification methods.

\vspace{8mm}
\hrule
\vspace{8mm}

\textbf{Acknowledgements:} This research has been conducted using the UK Biobank Resource under Application Number 98032. The initial draft of this paper was generated by the Deep Research feature of ChatGPT, a generative artificial intelligence model. It was then edited by human authors with further input from artificial intelligence. The initial ChatGPT prompt is provided in the Supplementary Material, as is the initial draft manuscript prepared by the model.

\bibliographystyle{../../master/wileyj}
\bibliography{../../master/masterref}

\clearpage

\renewcommand{\thesection}{A\arabic{section}}
\renewcommand{\thesubsection}{A.\arabic{subsection}}
\renewcommand{\thetable}{A\arabic{table}}
\renewcommand{\thefigure}{A\arabic{figure}}
\renewcommand{\theequation}{A\arabic{equation}}
\renewcommand{\thepage}{S\arabic{page}}
\setcounter{table}{0}
\setcounter{figure}{0}
\setcounter{equation}{0}
\setcounter{page}{1}
\renewcommand{\tablename}{Supplementary Table}
\renewcommand{\figurename}{Supplementary Figure}
\setcounter{section}{0}
\setcounter{subsection}{0}
\subsection*{Supplementary Material}
\subsection{Artificial intelligence prompt}
The prompt to write the initial draft of this paper was as follows: ``Could you please write a draft of a scientific paper that proposes a novel paradigm for Mendelian randomization by exploiting regional variation in the distribution of the risk factor of interest? The idea would be to perform separate Mendelian randomization analyses in different study populations that have different average levels of the exposure, and to use this to explore heterogeneity (and potentially, non-linearity) in effect estimates. The draft should have an Abstract, Introduction, Methods, Applied example, Results, and Discussion sections. The applied example will investigate the effect of vitamin D supplementation on coronary artery disease risk in different UK Biobank recruitment centres. There is no need to perform these analyses, but please leave space for us to include those analyses in the manuscript. Please clarify how this approach differs from other GxE methods, such as MR-Genius. The paper should be formatted for Statistics in Medicine, and aimed at an audience of applied statisticians and quantitative epidemiological researchers. The method could be called "dynamic contextual Mendelian randomization", but if you have a better name then we are happy to consider. Please provide an editable manuscript file, and code (in the R programming language) for running simulations to include in the submission (if you include simulations as part of the manuscript). We will read and edit the draft thoroughly before submission. Thank you for your help.''

\clearpage

\subsection{Simulation study code}
We here report the simulation study code, implemented using the R statistical software language:

\begin{verbatim}
library(meta)
library(RadialMR)
library(MendelianRandomization)

scens = 3
times = 1000
parts = 10000
groups = 10

set.seed(496)

alpha = rep(seq(from=8, by=0.2, length.out=groups), each = parts)
 # larger differences
alpha = rep(seq(from=9, by=0.1, length.out=groups), each = parts)
 # smaller differences
subgroup = rep(1:10, each = parts)
q1.val = NULL; q1.pval = NULL; qq1.val = NULL;
qq1.pval = NULL; trend1.pval = NULL;
q2.val = NULL; q2.pval = NULL; qq2.val = NULL;
qq2.pval = NULL; trend2.pval = NULL;
q3.val = NULL; q3.pval = NULL; qq3.val = NULL;
qq3.pval = NULL; trend3.pval = NULL;

 for (j in 1:times) {
    bx = NULL; by1 = NULL; by1se = NULL;
  bxse = NULL; by2 = NULL; by2se = NULL;
 xmean = NULL; by3 = NULL; by3se = NULL;
     g = rbinom(parts*groups, 2, 0.3)
     u = rnorm(parts*groups)
  epsx = rnorm(parts*groups)
 epsy1 = rnorm(parts*groups)
 epsy2 = rnorm(parts*groups)
 epsy3 = rnorm(parts*groups)
     x = alpha + 0.5*g + u + epsx
    y1 = 0.8*x - u + epsy1
    y2 = 0.04*x^2 - u + epsy2
    y3 = 0.25*ifelse(x>10, 1, 0)*(x-10) - u + epsy3
 for (k in 1:groups) {
 xmean[k] = mean(x[subgroup==k])
   bx.reg = summary(lm(x[subgroup==k]~g[subgroup==k]))
    bx[k] = bx.reg$coef[2]
  bxse[k] = bx.reg$coef[2,2]
  by1.reg = summary(lm(y1[subgroup==k]~g[subgroup==k]))
   by1[k] = by1.reg$coef[2]
 by1se[k] = by1.reg$coef[2,2]
  by2.reg = summary(lm(y2[subgroup==k]~g[subgroup==k]))
   by2[k] = by2.reg$coef[2]
 by2se[k] = by2.reg$coef[2,2]
  by3.reg = summary(lm(y3[subgroup==k]~g[subgroup==k]))
   by3[k] = by3.reg$coef[2]
 by3se[k] = by3.reg$coef[2,2]
   }

           est1 = sum(by1*bx*by1se^-2)/sum(bx^2*by1se^-2)
      q1.val[j] = sum( (by1/bx-est1)^2*(by1se/bx)^-2 )
           reg1 = metagen(by1/bx, by1se/bx)
          mreg1 = metareg(reg1, ~xmean)
     q1.pval[j] = reg1$pval.Q
 trend1.pval[j] = mreg1$pval[2]
           rad1 = ivw_radial(mr_input(bx, bxse, by1, by1se), weights=3,
                             alpha=1e-100, summary=FALSE)
     qq1.val[j] = rad1$q
    qq1.pval[j] = pchisq(rad1$q, df=groups-1, lower.tail=FALSE)

   est2 = sum(by2*bx*by2se^-2)/sum(bx^2*by2se^-2)
 q2.val[j] = sum( (by2/bx-est2)^2*(by2se/bx)^-2 )
   reg2 = metagen(by2/bx, by2se/bx)
  mreg2 = metareg(reg2, ~xmean)
 q2.pval[j] = reg2$pval.Q
 trend2.pval[j] = mreg2$pval[2]
           rad2 = ivw_radial(mr_input(bx, bxse, by2, by2se), weights=3,
                             alpha=1e-100, summary=FALSE)
     qq2.val[j] = rad2$q
    qq2.pval[j] = pchisq(rad2$q, df=groups-1, lower.tail=FALSE)

   est3 = sum(by3*bx*by3se^-2)/sum(bx^2*by3se^-2)
 q3.val[j] = sum( (by3/bx-est3)^2*(by3se/bx)^-2 )
   reg3 = metagen(by3/bx, by3se/bx)
  mreg3 = metareg(reg3, ~xmean)
 q3.pval[j] = reg3$pval.Q
 trend3.pval[j] = mreg3$pval[2]
           rad3 = ivw_radial(mr_input(bx, bxse, by3, by3se), weights=3,
                             alpha=1e-100, summary=FALSE)
     qq3.val[j] = rad3$q
    qq3.pval[j] = pchisq(rad3$q, df=groups-1, lower.tail=FALSE)
   }

sum(q1.pval<0.05)
sum(q2.pval<0.05)
sum(q3.pval<0.05)
sum(qq1.pval<0.05)
sum(qq2.pval<0.05)
sum(qq3.pval<0.05)
sum(trend1.pval<0.05)
sum(trend2.pval<0.05)
sum(trend3.pval<0.05)
\end{verbatim}

%
%

\end{document}